\documentclass{aastex}
\lefthead{van Belle}
\righthead{Predicting Stellar Angular Sizes}

\begin{document}
\title{Predicting Stellar Angular Sizes}
\author{Gerard T. van Belle}
\affil{Jet Propulsion Laboratory, California Institute of
Technology, \\4800 Oak Grove Drive, Pasadena, CA 91109
\\gerard@huey.jpl.nasa.gov}
\authoremail{gerard@huey.jpl.nasa.gov}

\begin{abstract}
Reliable prediction of stellar diameters, particularly angular
diameters, is a useful and necessary tool for the increasing
number of milliarcsecond resolution studies being carried out in
the astronomical community. A new and accurate technique of
predicting angular sizes is presented for main sequence stars,
giant and supergiant stars, and for more evolved sources such as
carbon stars and Mira variables. This technique uses observed $K$
and either $V$ or $B$ broad-band photometry to predict $V=0$ or
$B=0$ zero magnitude angular sizes, which are then readily
scaled to the apparent angular sizes with the $V$ or $B$
photometry.  The spread in the relationship is 2.2\% for main
sequence stars; for giant and supergiant stars, 11-12\%; and for
evolved sources, results are at the 20-26\% level.  Compared to
other simple predictions of angular size, such as linear
radius-distance methods or black-body estimates,
zero magnitude angular size
predictions can provide apparent angular sizes with errors that are 2 to 5
times
smaller.
\end{abstract}


\section{Introduction}

Prediction of stellar angular sizes is a tool that has come to be
used with greater frequency with the advent of high resolution
astronomical instrumentation.  Structure at the tens to hundreds
of milliarcsecond (mas) level is now being routinely observed with
the Hubble Space Telescope, speckle interferometry, and adaptive
optic systems.  Single milliarcsecond observations, selectively
available for many years with the technique of lunar occultations,
are now becoming less specialized as prototype interferometers in
the optical and near infrared evolve towards facility class
instruments.  For all of these telescopes and techniques, it is
often desirable to predict angular size of stars, to select either
appropriate targets or calibration sources.

Detailed photometric and spectrophotometric predictive methods
provide results with high accuracy (1-2\% diameters; cf.
\markcite{bl98}Blackwell \& Lynas-Gray 1998, \markcite{co96}Cohen
et al. 1996).  However, diameters from these methods require large
amounts of data that is often difficult to obtain, and as such,
are available for a limited number of objects. For the general
sample of stars, only limited information is available, and
spectral typing, photometry, and parallaxes are all less available
and less accurate as one examines stars at greater distances.
Deriving expected angular sizes is a greater challenge in this
case. Fortunately the general availability of $B$ or $V$ band
data, and forthcoming release of the data from the 2MASS and DENIS
surveys, which have limiting magnitudes of $K > 14.3$ and $13.5$,
respectively (\markcite{be98}Beichman et al. 1998,
\markcite{ep97}Epchtein 1997), will provide at least broad-band
photometry on these more distant sources. Given these databases,
of general utility is a method based strictly upon this widely
available data.  In this paper a method based solely upon $K$ and
either $B$ or $V$ broad-band photometry will be presented, and it
will be shown that angular sizes for a wide variety of sources can
be robustly predicted with merely two-color information. A similar
relationship is discussed by Mozurkewich et al. (1991), who
present a `distance normalized uniform disk angular diameter' as a
function of $R-I$ color, but with a limited number ($N=12$) of
objects to calibrate the relationship. Related to these methods is
the study of stellar surface brightness as a function of $V-K$
color published by \markcite{di93}Di Benedetto (1993), which built
on the previous work by Barnes \& Evans (\markcite{ba76a}Barnes \&
Evans 1976, \markcite{ba76b}Barnes et al. 1976,
\markcite{ba78}1978).

\section{Sources of Data\label{sec2sources}}

The relationship between angular size and color to be presented in
\S\ref{sec3_VK} is strictly empirical.  The angular sizes and
photometry utilized to calibrate the method are all available in
the literature, and in many cases are also online, and their
sources are presented below.

\subsection{Available Angular Size Data\label{sec3_lit_ref}}

As a test of the method, we shall be examining its
predictions against known angular
diameters.  For stars that have evolved off of the main sequence,
angular diameters as determined in the near-infrared are
preferred, as limb darkening - and the need for models to
compensate for it - is less than at shorter wavelengths.  There
are four primary sources in the literature of near-infrared
angular diameters (primarily K band):

\textit{Kitt Peak.} The lunar occultation papers by Ridgway and
his coworkers
(\markcite{ri74a}Ridgway et al. 1974,
\markcite{ri77b}Ridgway 1977,
\markcite{ri77a}\markcite{ri79}\markcite{ri80a}\markcite{ri80b}\markcite{ri82a}\markcite{ri82b}\markcite{ri82c}Ridgway et al. 1977, 1979, 1980a, 1980b, 1982a, 1982b, 1982c,
\markcite{sc86}Schmidtke et al. 1986) established the field of
measuring angular sizes of cool stars in the near-infrared. This
effort is no longer active.

\textit{TIRGO.}  The lunar occultation papers by Richichi and his
coworkers
(\markcite{ri88}\markcite{ri91}\markcite{92a}\markcite{92b}\markcite{ri95}\markcite{ri98a}\markcite{ri98b}\markcite{98c}\markcite{ri98d}\markcite{di91}Richichi
et al. 1988, 1991, 1992a, 1992b, 1995, 1998a, 1998b, 1998c, 1999,
Di Giacomo et al. 1991) have further developed this particular
technique of diameter determinations. The group is continuing to
explore the high-resolution data obtainable from lunar
occultations.  The recent publications from the TIRGO group
include data from medium to large aperture telescopes (1.23m -
3.5m), along with concurrent photometry.

\textit{IOTA.}  The K band angular diameters papers from the
Infrared-Optical Telescope Array by Dyck and his coworkers
(\markcite{dy96a}\markcite{dy96b}\markcite{dy98}Dyck et al. 1996a, 1996b, 1998,
\markcite{va96}\markcite{va97}\markcite{va99b}van Belle et al. 1996, 1997, 1999b) provided
a body of information on normal giant and supergiant papers, and
also on more evolved sources such as carbon stars and Mira
variables. Recently, results from this interferometer using the
FLUOR instrument have become available (\markcite{pe98}Perrin et
al. 1998).

\textit{PTI.}  Although there is only one angular diameter paper
currently available from the Palomar Testbed Interferometer
(\markcite{va99a}van Belle et al. 1999a), 69 objects are presented
in the manuscript from this highly automated instrument.

Altogether, this collection from the literature represents 92
angular diameters for 67 carbon stars and Miras, and 197 angular
diameters for 190 giant and supergiant stars.  In addition to
these near-infrared observations of evolved objects,
shorter wavelength observations
were used to obtain diameters for main sequence objects -- few
near infrared observations exist for these smaller sources.  These
objects were culled from the catalog by \markcite{fr88}Fracassini
et al. (1988), limiting the investigation to direct angular size
measures found in that catalog: lunar occultations, eclipsing and
spectroscopic binaries, and the intensity interferometer
observations of \markcite{ha74}Hanbury Brown et al. (1974).
Unfortunately, this sample of 50 main sequence objects is much
smaller than the evolved star sample, largely reflecting the
current resolution limits of roughly 1 mas in both the
interferometric and lunar occultation approaches: a one solar
radius object at a distance of 10 pc has an angular size of 0.92 mas.
Furthermore,
many of main sequence stars did not have sufficient photometry to
be used in the technique discussed in \S\ref{sec3_VK}.
Fortunately, added to this sample are the well-calibrated
measurements for the Sun (\markcite{al73}Allen 1973).

Shorter wavelength observations of giant and supergiant stars,
while available (eg., \markcite{hu89}Hutter et al. 1989,
\markcite{mo91}Mozurkewich et al. 1991), were not utilized in this
study for two reasons. First, there are complications arising from
reconciling angular diameters inferred from short wavelength
($\lambda < 1.2\mu$m) observations with the desired Rosseland mean
diameters for these cooler stars. Second, the majority of the data
collected on these stars, represented in the Mark III
interferometer database, remains unpublished. Fortunately, these
data are anticipated to be published soon
(\markcite{mo99}Mozurkewich 1999) and will be complimented by
additional short-wavelength data from the NPOI interferometer
(\markcite{no99}Nordgren 1999).

\subsection{Sources of Photometry
\label{sec3_bolo_sources}}

The widespread availability of Internet access, coupled with the
electronic availability of most (if not all) of the photometric
catalogs, has made task of researching archival photometry much
more tractable.  When photometry was not directly available from the
telescope observations, the archival sources utilized in this
investigation were as follows:

\textit{General Data.}  One of the more thorough references on
stellar objects is SIMBAD (\markcite{eg91}Egret et al. 1991;
http://simbad.u-strasbg.fr/ (France)
and http://simbad.harvard.edu/ (US Mirror)).  In addition to the
web-based query forms, one may also obtain information from SIMBAD
by telnet and email.  It is important to note that SIMBAD is
merely a clearing house of information from a wide variety of
sources and is not an original source in and of itself; any
information that ends up being crucial to the merit of an
astrophysical investigation should be checked against its primary
source.

\textit{Infrared Photometry ($\lambda>1\mu$m).}  The Catalog of
Infrared Observations (CIO), a extensive collection of IR
photometry by \markcite{ge93}Gezari et al. (1993) has been
updated, although the most recent version is available only online
(\markcite{ge97}Gezari, Pitts \& Schmitz 1997). The latter catalog
can be queried with individual stars or lists of objects at VizieR
(\markcite{ge97}Genova et al. 1997;
http://vizier.u-strasbg.fr/ (France) and
http://adc.gsfc.nasa.gov/viz-bin/VizieR (US Mirror)). As with the
SIMBAD data, the CIO is merely a collection of the data in the
literature, and examination of the primary sources is advised.
Also, as noted in the introduction, the forthcoming release of the 2MASS
and DENIS catalogs will greatly augment the collective database of
near-infrared photometry (whose home pages are
http://www.ipac.caltech.edu/2mass/ and
http://www-denis.iap.fr/denis.html, respectively).

\textit{Visual Photometry.} The General Catalog of Photometric
Data (GCPD) provides a large variety of wide- to narrow-band
visual photometric catalogs (Mermilliod et al. 1997;
http://obswww.unige.ch/gcpd/gcpd.html). For variable stars, the
American Association of Variable Star Observers (AAVSO) and its
French counterpart, the Association Fran\c{c}aise des Observateurs
d'Etoiles Variables (AFOEV) are both excellent sources of
epoch-specific visible light photometry (Percy \& Mattei 1993,
Gunther 1996;
http://www.aavso.org/
and http://cdsweb.u-strasbg.fr/afoev/, respectively).

\section{Zero Magnitude Angular Size versus $V-K$, $B-K$ Colors \label{sec3_VK}}

The large body of angular sizes now available allows for direct
predictions of expected angular sizes, bypassing many
astrophysical considerations, such as atmospheric structure,
distance, spectral type, reddening, and linear size. To compare
angular sizes of stars at different distances, one approach is to
scale the sizes relative to a zero magnitude of $V=0$:
\begin{equation}
\theta_{V=0} = \theta \times 10^{V/5}.
\end{equation}
The angular size thus
becomes a measure of apparent surface brightness (a
more detailed discussion of related quantities
may be found in \markcite{di93}Di Benedetto 1993.)  Conversion
between a $V=0$ zero magnitude angular size, $\theta_{V=0}$,
and actual angular
size, $\theta$, is trivial with a known $V$
magnitude and the equation above.
The same approach has been employed for $K=0$ (see
\markcite{dy96a}Dyck et al. 1996a) and will also be applied in this
paper to $B=0$. Given the general prevalence of $V$ band and the
inclusion of $B$ band data in the 2MASS catalog, the apparent
angular size approach will be developed here for $V-K$ and $B-K$
colors.

\subsection{Evolved Sources: Giant and Supergiant Stars\label{sec_giantVK}}

163 normal giant and supergiant stars found in the interferometry
and lunar occultation papers were also found to have available $V$
photometry. By examining their near-infrared angular sizes, we can
establish a relationship between $V=0$ zero magnitude angular
size and $V-K$ color:
\begin{equation}
\theta_{V=0} = 10^{0.669\pm0.052 + 0.223\pm0.010 \times(V-K)}.
\label{eqn_VKgiant}
\end{equation}
The errors on the 2 parameters in the equation above are $1\sigma$
errors determined from a $\chi^2$ minimization; given 2 degrees of
freedom in the equation, $\Delta \chi^2=2.30$ about the $\chi^2$
minimum for this case (\markcite{pr92}Press et al. 1992). Similar
error calculations will be given for all other relationships
reported in this manuscript.  Examining the distribution of the
differences between the fit and the measured values,
$\Delta\theta_{V=0}$, we find an approximately Gaussian
distribution with the rms value of the 163 differences yielding a
fractional error of ${(\Delta\theta_{V=0} / \theta_{V=0})_{rms}}=
11.7\%$.

Similarly, for $B-K$ color, 136 giant and supergiant stars had
available photometry, resulting in the following fit:
\begin{equation}
\theta_{B=0} = 10^{0.648\pm0.072 + 0.220\pm0.012 \times(B-K)},
\label{eqn_BKgiant}
\end{equation}
with an rms error of 10.8\%.

\begin{figure}[t]
\plotone{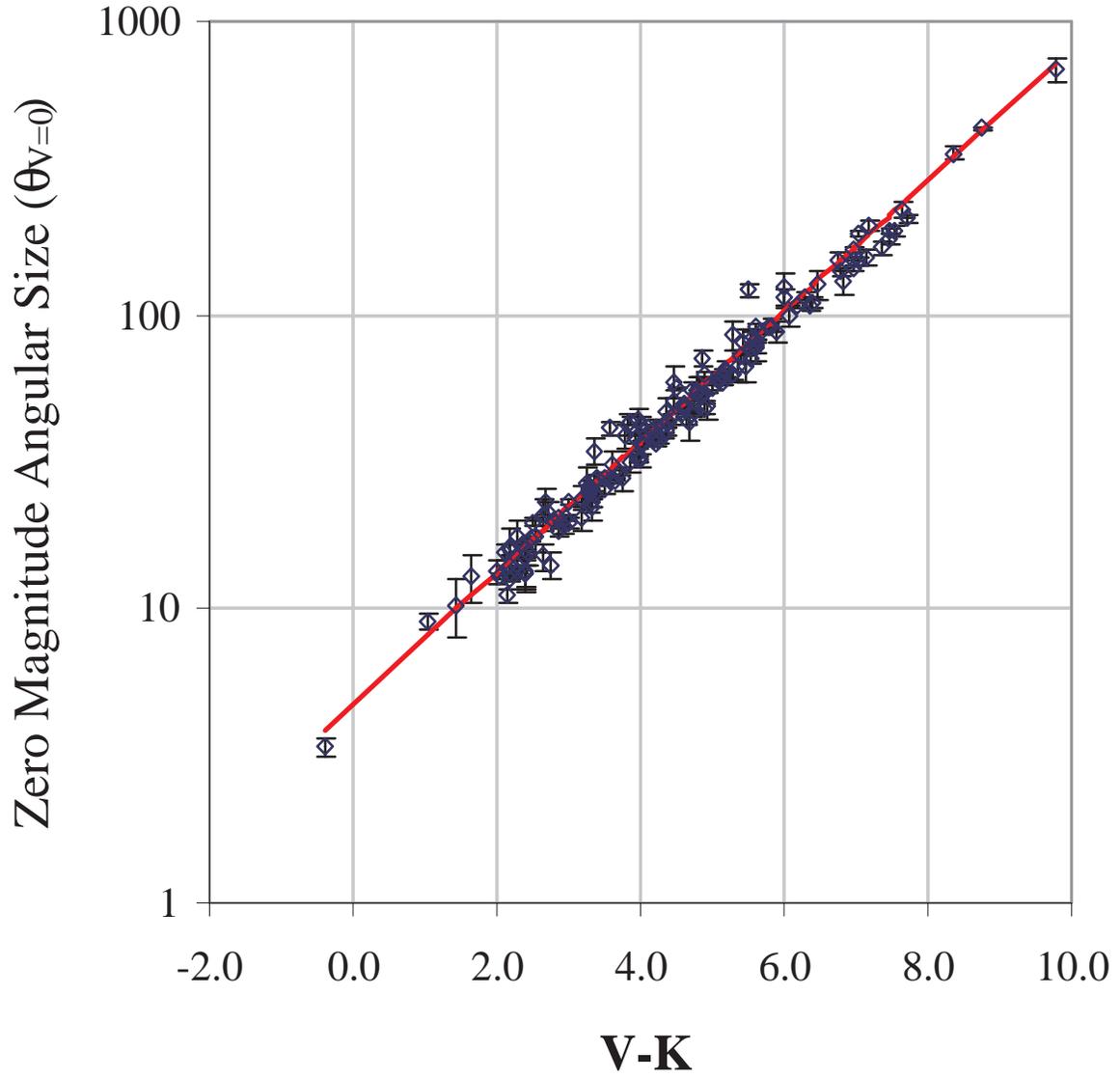} \caption{The $\theta_{V=0}$ zero magnitude angular
size versus $V-K$ color for luminosity class I, II, and III giant
stars.} \label{fig2}
\end{figure}

\begin{figure}[t]
\plotone{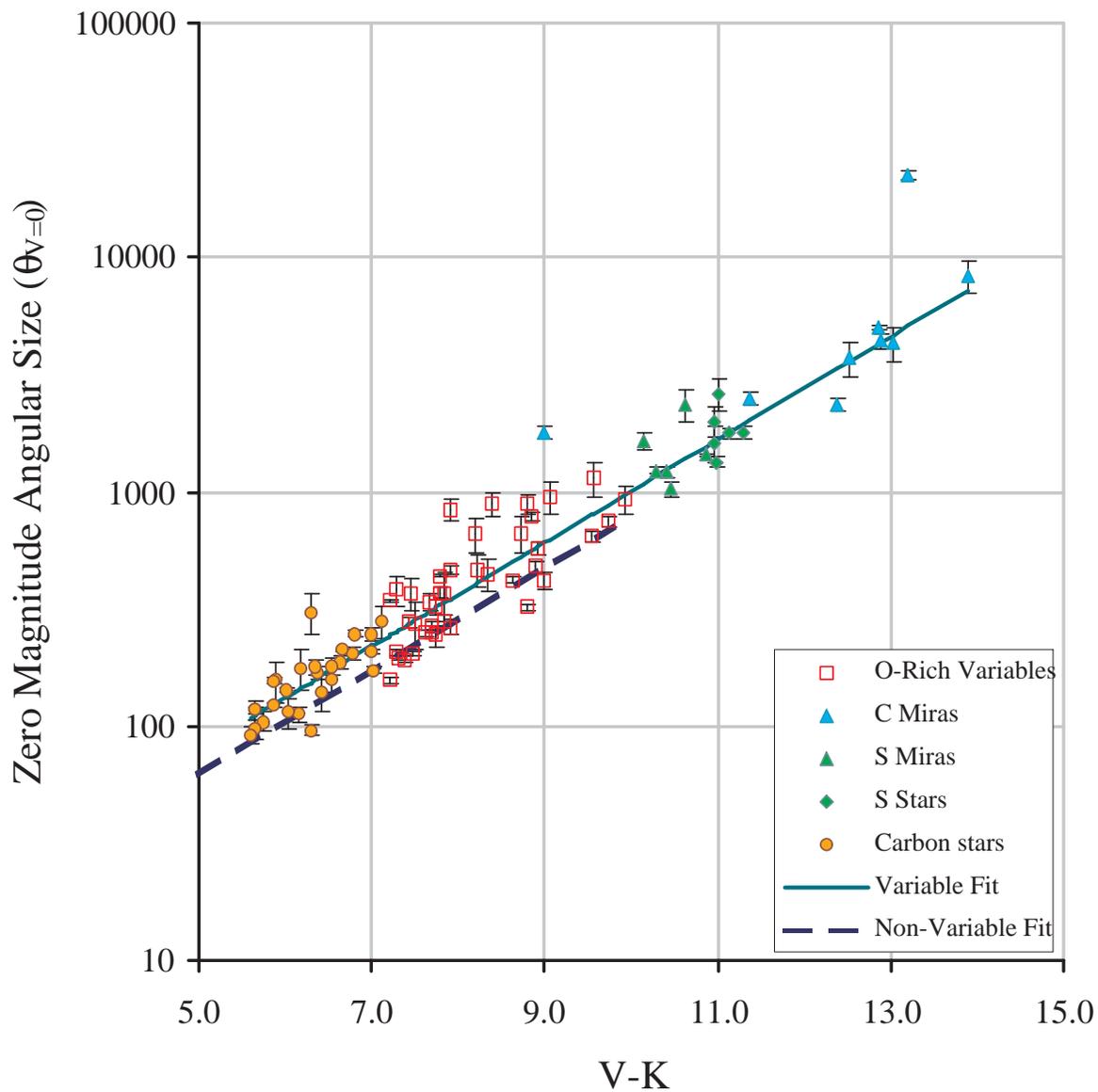} \caption{The $\theta_{V=0}$ zero magnitude angular
size versus $V-K$ color for evolved stars, including carbon stars,
S stars, all types of Mira variables, and non-Mira variables. The
solid upper line is the fit line for these objects, and the dashed
lower line is the fit line for the giants and supergiants from
Figure \ref{fig2}.} \label{fig3}
\end{figure}

The relationship appears valid over a $V-K$ range of 2.0 to 8.0.
Blueward of $V-K=2.0$, the subsample is too small ($N=3$) to
confidently indicate whether or not the fit is valid, in spite of
the goodness of fit for the whole subsample.  The same is true
redward of $V-K=8.0$.  Also, for stars redward of approximately
$V-K = 8$, care must be taken to exclude variable stars (both
semiregular and Miras).  The data points and the fit noted above
may be seen in Figure \ref{fig2}; $\theta_{V=0}$ and standard
deviation by $V-K$ bin is given in Table \ref{tab4}.  The Miras
are plotted separately in Figure \ref{fig3} and will be discussed
below.

\begin{figure}[t]
\plotone{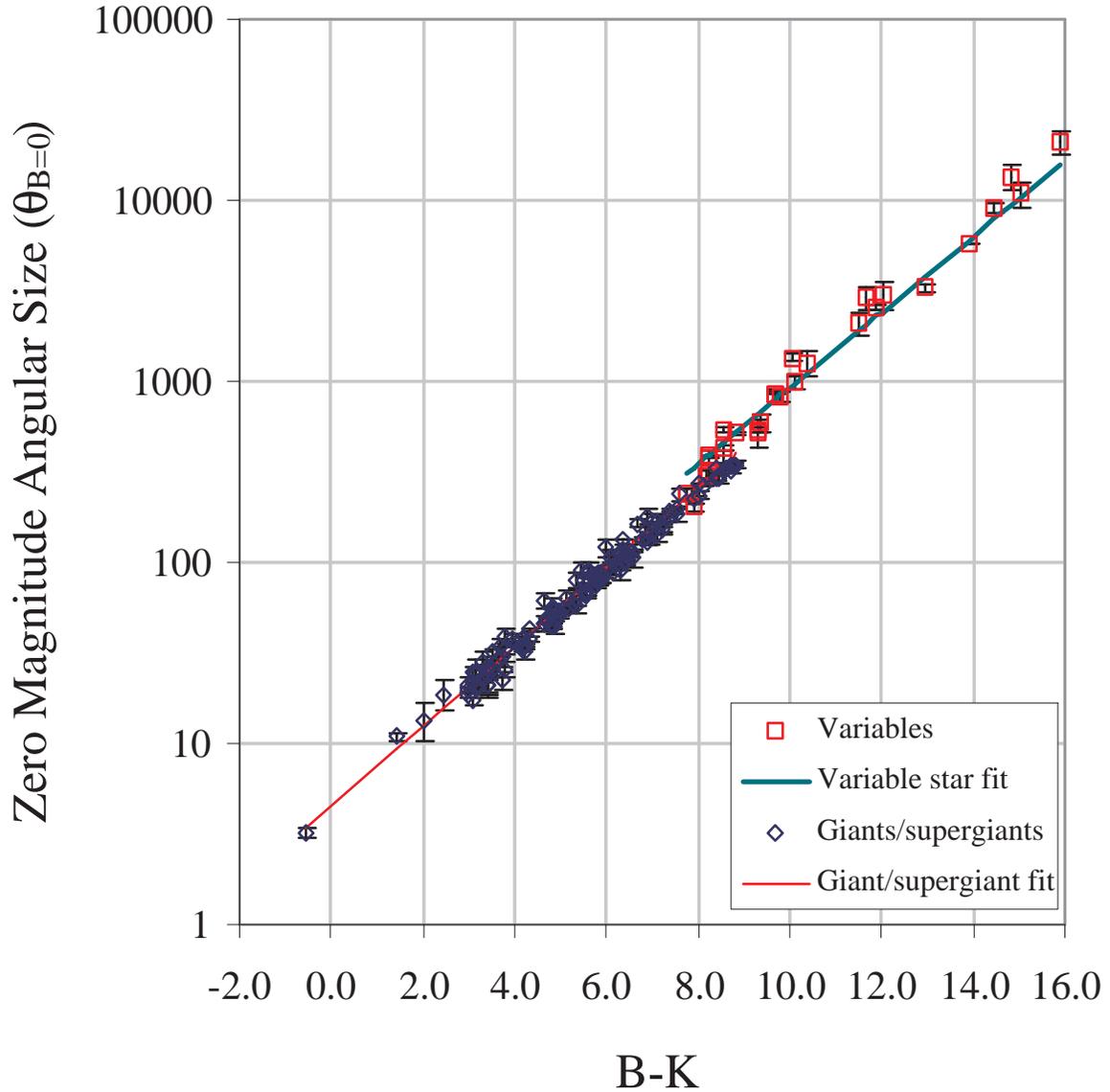} \caption{The $\theta_{B=0}$ zero magnitude angular
size versus $B-K$ color for giant/supergiant stars and evolved
stars, which includes Mira variables, S stars, carbon stars, and
non-Mira variables.  The upper line is the fit line for the
evolved stars, the lower line is the fit line for the giants and
supergiants.} \label{fig4}
\end{figure}

For $B-K$ between 3.0 and 7.5, the relationship exhibits a
similar if not slightly superior validity.  As with the $V-K$
color, the relationship appears to be valid down blueward of the
short edge of that range, down to $B-K=-1$, but the data are
sparse. Redward of $B-K=7.5$, the relationship also exhibits
potential confusion with the Mira variable stars, although there
appears to be less degeneracy, but this is possibly due to a
lesser availability of $B$ band data on these very red sources.
The data points and the fit noted above may be seen in Figure
\ref{fig4}; $\theta_{B=0}$ and standard deviation by $B-K$ bin is
given in Table \ref{tab4a}.

The potential misclassification of more evolved sources such as
carbon stars and variables (Miras or otherwise) as normal giant
and supergiant stars is a significant secondary consideration. For
the dimmer sources for which little data is available,
non-classification is perhaps the more appropriate term.  What is
reassuring with regards to the issue of classification errors is
that the robust relationships between $(\theta_{V=0},
V-K)$ and $(\theta_{B=0}, B-K)$ are valid for stars of luminosity
class I, II, and III, and that the more evolved stars occupy a
redder range of $B-K$ and $V-K$ colors (cf. \S\ref{var_stars}).
Since the $\theta_{V=0}$ and $\theta_{B=0}$ relationships
are insensitive to errors in luminosity class, this method is more
robust than the linear radius-distance method, particularly for
those stars in the $2.0 < V-K < 6.0$ and $3.0 < B-K < 7.5$ ranges,
where few if any stars of significant variability exist.  This
relationship is also considerably easier to employ than the method
of blackbody fits.

\subsection{Evolved Sources: Variable Stars\label{var_stars}}

By examining the 2.2
$\mu$m angular sizes for the 87 observations of 65 semiregular
variables, Mira variables and carbon stars (broadly classified
here as `variable stars') found in the literature, we can
establish a relationship between $V=0$ zero magnitude angular size and
$V-K$ color:
\begin{equation}
\theta_{V=0} = 10^{0.789\pm0.119 + 0.218\pm0.014\times(V-K)}.
\end{equation}
The rms error associated with this fit is 26\%.  The data points and
the fit noted above may be seen in Figure 3.  Similarly, for $B-K$
color, 19 evolved sources had available photometry for 29 angular
size observations, resulting in the following fit:
\begin{equation}
\theta_{B=0} = 10^{0.840\pm0.096 + 0.211\pm0.008\times(B-K)},
\end{equation}
with a rms error of 20\%.

For the variable stars, the relationship appears valid over
$V-K$, $B-K$ ranges of 5.5 to 13.0 and 9.0 to 16.0, respectively.
Redward of $V-K=13$, the sample is too small ($N=3$) to
confidently indicate whether or not the fit is valid, in spite of
the goodness of fit for the general sample.  It is interesting to
note that the slope of the fits for the variable stars and for the
giant/supergiant stars is statistically identical for both $V-K$
and $B-K$ colors; only the intercepts are different.  This
corresponds to a $\theta_{V=0}$ size factor of $1.40\pm0.15$
between the smaller normal and and the larger variable stars
for a given $V-K$ color, and a corresponding
$\theta_{B=0}$ size factor of $1.34\pm0.21$.

\subsection{Main Sequence Stars}

\begin{figure}[t]
\plotone{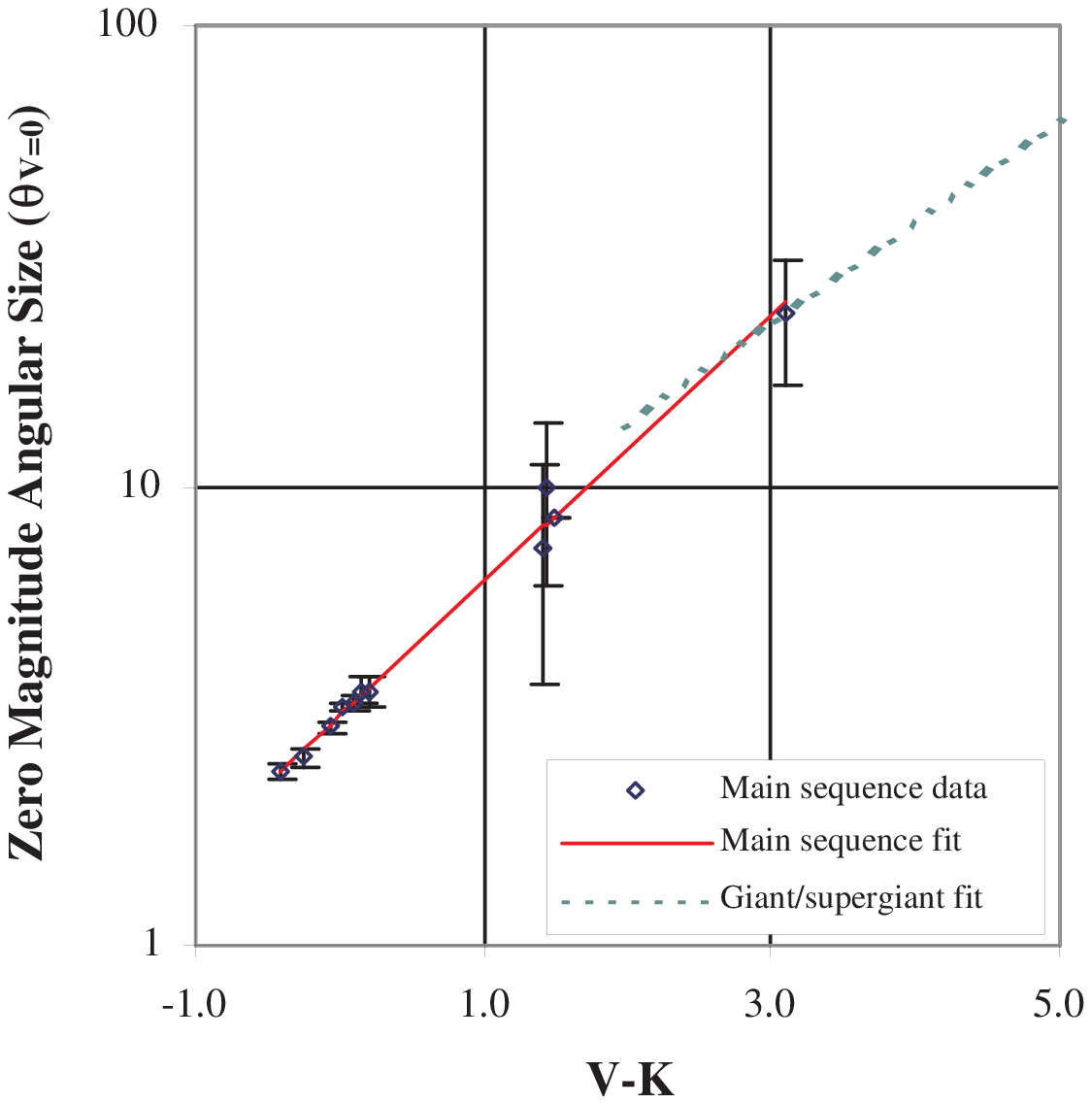} \caption{The $\theta_{V=0}$ zero magnitude angular
size versus $V-K$ color for main sequence stars.  The diamonds and
solid line are the data points and fit for B, A, and G type stars,
respectively; the dotted line is the fit for luminosity class I, II, and III
stars.} \label{fig5}
\end{figure}

By examining the objects in the
\markcite{fr88}Fracassini catalog (1988; specifically, many
objects from \markcite{ha74}Hanbury Brown et al. 1974), there
appears to be similar relationships between the $V-K$ \& $B-K$
colors, and $\theta_{V=0}$ \& $\theta_{B=0}$ angular sizes.
The sample set
of stars with adequate photometry is unfortunately limited to 11
objects.  However, the Hanbury Brown objects and the Sun are
measured with high accuracy and allow for accurate calibration of
the stellar zero magnitude angular sizes
in the ranges of $-0.4 < V-K < +1.5$ and $-0.6 < B-K <
+2.0$.
Limiting the fit analysis to the robust measurements from
Hanbury Brown and for the sun, the relationships between the
colors and their zero magnitude angular sizes are
\begin{eqnarray}
\theta_{V=0} = 10^{0.500\pm0.023 + 0.264\pm0.012
\times(V-K)}\textrm{,}&\textrm{and}\\
\theta_{B=0} = 10^{0.500\pm0.012 + 0.290\pm0.016 \times(B-K)}.&
\end{eqnarray}
The resulting rms errors are only 2.2\% for both the $V-K$ and
$B-K$ relationships. The $\theta_{V=0}$ versus $V-K$ data
for these objects are
plotted in Figure \ref{fig5}; the $\theta_{B=0}$ versus $B-K$ data are similar
in appearance and will not be plotted. The relationship holds not only
for the B and A type objects in the $-0.5 < V-K <
+0.5$ range, but also for the Sun at $V-K
\approx 1.5$.  Also plotted is the fit for giants and supergiants,
which has a slightly different slope; the two fits are shown intersecting
at $V-K
\approx 2.5$, although due to poor sampling in this region it is unclear
how (or if) the two functions truly join.

\subsection{Analysis of Errors}

As was given in \S\ref{sec_giantVK}, the rms fractional error
between the measured and predicted values for $\theta_{V=0}$
versus $V-K$ for giants and supergiants is ${(\Delta\theta_{V=0} /
\theta_{V=0})_{rms}}= 11.7\%$. There are three components of this
error: (1) Angular size errors, (2) Errors in $V-K$, and (3)
Deviations in the relationship due to unparameterized phenomena,
which shall be broadly labeled `natural dispersion' in the
relationship and will be discussed in more detail below. For the
first component, the rms fractional error of the 163 measured
$\theta$ values found in the literature is $(\Delta\theta /
\theta)_{rms} = 6.9\%$. For the photometry, given the
heterogeneous sources, we estimate that the $V$ and $K$ photometry
will have errors between 0.1 and 0.2 magnitudes (resulting in
$V-K$ color errors of 0.14 to 0.28 magnitudes), which would result
in an size error of 3.1-6.3\%. Finally, subtracting these two
sources of measurement error in quadrature from the measured
dispersion, a natural dispersion in the relationship between 7.0
and 8.9\% remains. A similar analysis for the giant/supergiant
$\theta_{B=0}$ versus $B-K$ results in $(\Delta\theta /
\theta)_{rms} = 7.0\%$ for the 136 observations, indicating of
5.2-7.6\% of natural dispersion.  For both of these relationships,
the natural dispersion is a factor as significant as the errors in
angular size, and potentially the dominant factor.

For the main sequence stars, the errors in angular size for both
colors were $(\Delta\theta / \theta)_{rms} = 4.5\%$; the
errors in photometry were expected to be no different than the giant/supergiant
stars, at 0.1-0.2 magnitudes per photometric band.
The main sequence stars exhibited no measureable levels of natural
dispersion,
being able
to fully account for the observed rms spread in both the $V-K$ and
$B-K$ relationships with angular size or color errors.

For the variable stars, the difficulties in
obtaining contemporaneous photometry result in larger measurement
error, despite steps taken to ensure epoch-dependent observations. As such,
the errors are expected to be between 0.2 and 0.4 magnitudes for
the individual $V$ and $K$ measurements. The resulting
characterization of natural dispersion of 20-23\% for the $V-K$
relationship, and 12-16\% for $B-K$, dominating the
angular size dispersion of $(\Delta\theta / \theta)_{rms} = 10\%$
for both colors.

The specific nature of the natural dispersion term in the rms
error is potentially due to stellar surface properties that affect
current one-dimensional angular size determination techniques. The
limited observations of individual objects with two-dimensional
and more complete spatial frequency coverage have indicated
asymmetries in stellar atmospheres that could potentially affect
size determinations from both interferometric and lunar
occultations. Early measurements of this nature were detection of
asymmetries in the envelope of $o$ Cet with speckle interferometry
(\markcite{ka91}Karovska et al. 1991).  Direct imaging of the
surface of $\alpha$ Ori has provided evidence of a large hot spot
on that supergiant's surface (\markcite{gi96}Gilliland \& Dupree
1996). More recently, similar evidence for aspheric shapes of
other Miras has been obtained, also with HST
(\markcite{la97}Lattanzi et al. 1997), and evidence for more
complicated morphologies in the structure of the M5 supergiant VY
CMa in the near-IR has been obtained using nonredundant aperture
masking on Keck 1 (\markcite{mo99}Monnier et al. 1999, \markcite{tu99}Tuthill et al. 1999). Various
atmospheric phenomena, such as nonradial pulsations, spots on the
stellar surface, and rotational distortion of the stellar
envelope, potentially explain these observations. The progressive
increase along stellar evolutionary states in observed natural
dispersion from undetectable levels with the main sequence stars
to dominant levels with the most evolved sources is consistent
with the onset of these phenomena more significantly associated
with extended atmospheres.

{\it Interstellar Extinction.}  A brief discussion of the
potential impact of interstellar extinction upon the results
presented herein is warranted.  The empirical reddening
determination made by \markcite{ma80}Mathis (1980), which agrees very well with
van de Hulst's theoretical reddening curve number 15 (see \markcite{jo68}Johnson
1968), predicts that $A_K = 0.11 A_V$.  From that relative
reddening value, the effect of interstellar reddening upon the
various angular size expressions may be derived to be:
\begin{eqnarray}
{\theta_{V=0}}' = \theta_{V=0}\times 10^{0.225\times[ {(V-K)}' -
(V-K)]}\textrm{,}&\textrm{ and}\label{eqn_VKred}
\\
{\theta_{V=0}}' = \theta_{V=0}\times 10^{0.218\times[
{(V-K)}' - (V-K)]}.\label{eqn_BKred}
&
\end{eqnarray}
Comparison of the slopes of equations (\ref{eqn_VKgiant}) and
(\ref{eqn_BKgiant})
with (\ref{eqn_VKred}) and (\ref{eqn_BKred}) demonstrates that the
angular size predictions for giant and supergiant stars are almost wholly unaffected by the
effects of interstellar extinction: any apparent reddening of a star's $V-K$ or $B-K$ color is
accompanied by an increase in the associated zero magnitude angular
size, along the slope of the predict lines.  This effect is
independent of the absolute amount of reddening encountered by a star,
since it is a {\it relative} effect between the two bandpasses of a given
color.

For main sequence stars, the slopes between the predict lines and reddening effect
indicates a gradual underestimation in actual stellar size as reddening increases.
Based upon typical reddening values of $A_V=0.8-1.9$ mag/kpc, a 2.2\%
effect (consistent with the expected level of error in the angular
size prediction)
will be present for stars with $A_V=0.18$, corresponding to  distances between 95 and 225 pc.  For
the variable stars, the slopes also skew, but slightly less so,
and in the opposite sense: the gradual trend will be to
overestimate sizes for reddened sources.  A 20\% effect for these
stars will be present for stars at $A_V=16.9$, corresponding to distances between 8.8 and 21
kpc - clearly not a significant factor for the current accuracy of
either the $V-K$ or $B-K$ relationship.

\section{Comparison of the Various Methods}

Previous approaches for estimating angular sizes have included
estimates of stellar linear size coupled with distance
measurements or estimates, and the extraction of angular sizes by
treating the objects as blackbody radiators.  The release of the
Hipparcos catalog (\markcite{pe97}Perryman et al. 1997), with its
parallax data, has increased the utility of the first method.
Spectral type and $V-K$ color have been explored as indicators of
intrinsic linear size for giant stars (\markcite{va99a}van Belle
et al. 1999a). Similarly, spectral type can be used to predict
linear size for main sequence stars (\markcite{al73}Allen 1973),
although this relationship appears to be poorly characterized.
There does not appear to be a $V-K$-linear radius relationship
presented for these stars in the literature, which would be
consistent with both photometric bands being on the Rayleigh-Jeans
tail of the blackbody curve for these hotter ($T>6000K$) objects.
The relative errors for predicting stellar angular diameters were
calculated as discussed in \S\ref{sec3_VK} for the stars in the
\S\ref{sec2sources} sample using these alternative methods, and
are summarized in Table \ref{tab5}.  For all of the stars in
question, deriving an apparent angular size from a $\theta_{V=0}$
or $\theta_{B=0}$ zero magnitude angular size delivers the
best results.

\section{Conclusion}

The new approach of establishing the $\theta_{V=0}$ and
$\theta_{B=0}$ zero magnitude angular
sizes appears to be an unrecognized yet
powerful tool for predicting the apparent angular sizes of stars of all
classes.  The very modest data requirements of this method make
it an ideal tool for quantification of this fundamental stellar
parameter.

\acknowledgments

Part of the work described in this paper was performed at the Jet
Propulsion Laboratory, California Institute of Technology under
contract with the National Aeronautics and Space Administration. I
would like to thank Andy Boden, Mark Colavita, Mel Dyck, Steve
Ridgway, and Bob Thompson for thoughtful comments during the
development of this manuscript, and an anonymous referee who provided
valuable feedback during the publication process.
This research has made use of the
SIMBAD, VizieR, and AFOEV databases, operated by the CDS,
Strasbourg, France. In this research, we have used, and
acknowledge with thanks, data from the AAVSO International
Database, based on observations submitted to the AAVSO by variable
star observers worldwide.

\begin{deluxetable}{cccccccccccc}
\scriptsize \tablewidth{0pc} \tablecaption{Zero Magnitude Angular Size
for Giants and Supergiants
as a Function of $V-K$ Color\label{tab4}} \tablehead{ &
\multicolumn{4}{c}{Normal Giants and Supergiants} &&
\multicolumn{4}{c}{Variables} &
\\
\cline{2-5} \cline{7-10}\\ \colhead{$V-K$} &
 &
& \colhead{Std.} & && & & \colhead{Std.} & & \\ \colhead{Center} &
\colhead{$N$} & \colhead{$\overline{\theta}_{V=0}$} &
\colhead{Dev.} & \colhead{Fit} && \colhead{$N$} &
\colhead{$\overline{\theta}_{V=0}$} & \colhead{Dev.} &
\colhead{Fit} & \colhead{Ratio} } \startdata -0.5 & 1 & 3.4 &  &
3.7 && 0 &  &  &  &  \nl 0.0 & 0 &  &  & 4.8 && 0 &  &  &  &  \nl
0.5 & 0 &  &  & 6.2 && 0 &  &  &  &  \nl 1.0 & 1 & 9.1 &  & 8.0 &&
0 &  &  &  &  \nl 1.5 & 2 & 11.6 & 1.8 & 10.3 && 0 &  &  &  &  \nl
2.0 & 9 & 13.9 & 1.7 & 13.3 && 0 &  &  &  &  \nl 2.5 & 17 & 16.7 &
3.1 & 17.2 && 0 &  &  &  &  \nl 3.0 & 12 & 20.5 & 3.1 & 22.3 && 0
&  &  &  &  \nl 3.5 & 20 & 27.2 & 4.4 & 28.7 && 0 &  &  &  &  \nl
4.0 & 21 & 37.8 & 4.4 & 37.1 && 0 &  &  &  &  \nl 4.5 & 15 & 47.0
& 5.7 & 47.9 && 0 &  &  &  &  \nl 5.0 & 18 & 58.2 & 5.9 & 61.9 &&
0 &  &  &  &  \nl 5.5 & 15 & 80.3 & 13.9 & 79.9 && 4 & 105 & 13 &
103 & $1.31\pm0.28$ \nl 6.0 & 7 & 102.7 & 13.3 & 103.1 && 7 & 140
& 25 & 132 & $1.37\pm0.30$ \nl 6.5 & 5 & 122.9 & 18.3 & 133.1 && 9
& 181 & 57 & 170 & $1.47\pm0.51$ \nl 7.0 & 9 & 159.6 & 23.5 &
171.9 && 8 & 233 & 60 & 220 & $1.46\pm0.43$ \nl 7.5 & 6 & 197.0 &
21.0 & 221.9 && 14 & 270 & 62 & 283 & $1.37\pm0.35$ \nl 8.0 & 0 &
&  & 286.6 && 9 & 461 & 184 & 365 &  \nl 8.5 & 1 & 355.4 &  &
370.0 && 4 & 605 & 217 & 470 & $1.70\pm0.61$ \nl 9.0 & 1 & 431.0 &
& 477.7 && 7 & 631 & 245 & 605 & $1.46\pm0.57$ \nl 9.5 & 0 &  &  &
&& 3 & 841 & 259 & 780 &  \nl 10.0 & 0 &  &  &  && 2 & 1286 & 511
& 1005 &  \nl 10.5 & 0 &  &  &  && 4 & 1456 & 604 & 1295 &  \nl
11.0 & 0 &  &  &  && 6 & 1795 & 465 & 1669 &  \nl 11.5 & 0 &  &  &
&& 2 & 2146 & 498 & 2150 &  \nl 12.0 & 0 &  &  &  && 0 &  &  &  &
\nl 12.5 & 0 &  &  &  && 2 & 3033 & 965 & 3569 &  \nl 13.0 & 0 &
&  &  && 0 &  &  &  &  \nl 13.5 & 0 &  &  &  && 0 &  &  &  &  \nl
14.0 & 0 &  &  &  && 1 & 8323 &  & 7635 &  \nl \tablecomments{The
number of stars $N$, average size $\theta_{V=0}$, and standard
deviation for each bin is given for both normal giant and
supergiant stars, and for variables, inclusive of Miras,
semi-regulars, and carbon stars.  The fits given are those
discussed in \S \ref{sec3_VK}; the ratios given are the average
$\theta_{V=0}$ size ratios for those $V-K$ bins where values exist
for both giant/supergiant stars and variables.  In general, the
variable stars have a $\theta_{V=0}$ size that is $1.44\pm0.15$
larger than their `normal' star counter parts for a given $V-K$
color.}
\enddata
\end{deluxetable}

\begin{deluxetable}{ccccccccccc}
\scriptsize \tablewidth{0pc} \tablecaption{Zero Magnitude Angular Size
for Giants and Supergiants
as a Function of $B-K$ Color\label{tab4a}} \tablehead{ &
\multicolumn{4}{c}{Normal Giants and Supergiants} &&
\multicolumn{4}{c}{Variables} &
\\
\cline{2-5} \cline{7-10}\\ \colhead{$B-K$} &&& \colhead{Std.} &
&&&& \colhead{Std.}\\ \colhead{Bin Center} & \colhead{$N$} &
\colhead{$\overline{\theta}_{B=0}$} & \colhead{Dev.} &
\colhead{Fit} && \colhead{$N$} &
\colhead{$\overline{\theta}_{B=0}$} & \colhead{Dev.} &
\colhead{Fit} & \colhead{Ratio} } \startdata -0.5 & 1 & 3.2 &  &
3.5 && 0 &  &  &  &  \nl 0.0 & 0 &  &  & 4.5 && 0 &  &  &  &  \nl
0.5 & 0 &  &  & 5.8 && 0 &  &  &  &  \nl 1.0 & 0 &  &  & 7.5 && 0
&  &  &  &  \nl 1.5 & 1 & 10.9 &  & 9.7 && 0 &  &  &  &  \nl 2.0 &
1 & 13.6 &  & 12.5 && 0 &  &  &  &  \nl 2.5 & 1 & 18.7 &  & 16.1
&& 0 &  &  &  &  \nl 3.0 & 10 & 21.4 & 2.7 & 20.7 && 0 &  &  &  &
\nl 3.5 & 13 & 26.2 & 4.2 & 26.7 && 0 &  &  &  &  \nl 4.0 & 11 &
34.5 & 3.6 & 34.4 && 0 &  &  &  &  \nl 4.5 & 6 & 47.2 & 8.2 & 44.3
&& 0 &  &  &  &  \nl 5.0 & 15 & 51.9 & 5.3 & 57.1 && 0 &  &  &  &
\nl 5.5 & 14 & 74.9 & 11.5 & 73.6 && 0 &  &  &  &  \nl 6.0 & 18 &
89.5 & 12.0 & 94.8 && 0 &  &  &  &  \nl 6.5 & 12 & 114.4 & 19.3 &
122.1 && 0 &  &  &  &  \nl 7.0 & 13 & 151.5 & 15.9 & 157.4 && 0 &
&  &  &  \nl 7.5 & 6 & 196.1 & 21.3 & 202.8 && 0 &  &  &  &  \nl
8.0 & 4 & 248.4 & 23.2 & 261.2 && 6 & 304 & 75 & 352 &
$1.23\pm0.32$ \nl 8.5 & 7 & 315.6 & 21.8 & 336.6 && 3 & 451 & 84 &
447 & $1.43\pm0.28$ \nl 9.0 & 2 & 344.2 & 5.5 & 433.7 && 1 & 520 &
& 569 &  \nl 9.5 & 0 &  &  &  && 5 & 669 & 164 & 723 &  \nl 10.0 &
0 &  &  &  && 3 & 1057 & 273 & 919 &  \nl 10.5 & 0 &  &  &  && 1 &
1270 &  & 1169 &  \nl 11.0 & 0 &  &  &  && 0 &  &  &  &  \nl 11.5
& 0 &  &  &  && 2 & 2501 & 561 & 1889 &  \nl 12.0 & 0 &  &  &  &&
2 & 2802 & 316 & 2402 &  \nl 12.5 & 0 &  &  &  && 0 &  &  &  &
\nl 13.0 & 0 &  &  &  && 1 & 3302 &  & 3883 &  \nl 13.5 & 0 &  &
&  && 0 &  &  &  &  \nl 14.0 & 0 &  &  &  && 1 & 5797 &  & 6276 &
\nl 14.5 & 0 &  &  &  && 1 & 9077 &  & 7979 &  \nl 15.0 & 0 &  &
&  && 2 & 12161 & 1755 & 10144 &  \nl \tablecomments{The number of
stars $N$, average size $\theta_{B=0}$, and standard deviation for
each bin is given for both normal giant and supergiant stars, and
for variables, inclusive of Miras, semi-regulars, and carbon
stars.  The fits given are those discussed in \S \ref{sec3_VK};
the ratios given are the average $\theta_{B=0}$ size ratios for
those $B-K$ bins where values exist for both giant/supergiant
stars and variables.  In general, the variable stars have a
$\theta_{B=0}$ size that is $1.34\pm0.21$ larger than their
`normal' star counter parts for a given $B-K$ color.}
\enddata
\end{deluxetable}

\begin{deluxetable}{llcclll}
\footnotesize \tablewidth{0pc} \tablecaption{Comparison of the
Various Angular Size Prediction Methods\label{tab5}} \tablehead{
\multicolumn{2}{c}{Method} & & \colhead{Error}  & \colhead{Notes}
} \startdata
\sidehead{Main Sequence Stars}
 & Linear Radius by Spectral Type &  & 25\% \nl
 & Linear Radius by $V-K$ Color & & N/A \nl
 & Angular Size by BBR Fit &  & 13\%  \nl
 & $V=0$ Angular Size by $V-K$ Color &  & 2.2\%  &\nl
 & $B=0$ Angular Size by $B-K$ Color &  & 2.2\% & \nl
\sidehead{Giant, Supergiant Stars}
 & Linear Radius by Spectral Type &  & 22\% & Giants only\nl
 & Linear Radius by $V-K$ Color &  & 22\% & Giants only\nl
 & Angular Size by BBR Fit &  & 18\%  \nl
 & $V=0$ Angular Size by $V-K$ Color &  & 11.7\%  \nl
 & $B=0$ Angular Size by $B-K$ Color &  & 10.8\%  \nl
\sidehead{Variable Stars}
 & $V=0$ Angular Size by $V-K$ Color &  & 26\%  \nl
 & $B=0$ Angular Size by $B-K$ Color &  & 20\%  \nl
\tablecomments{Errors given above are percentage errors relative
to the value predicted by each method.}
\enddata
\end{deluxetable}

\begin{references}
\reference{al73} Allen, C.W. 1973, Astrophysical Quantities, (London: Athlone Press)
\reference{ba76a} Barnes, T.G., \& Evans, D.S. 1976, MNRAS, 174, 489
\reference{ba78} Barnes, T.G., Evans, D.S., \& Moffett, T.J. 1978, MNRAS, 183, 285
\reference{ba76b} Barnes, T.G., Evans, D.S., \& Parsons, S.B. 1976, MNRAS, 174, 503
\reference{be98} Beichman, C.A., Chester, T.J., Skrutskie, M., Low, F.J., \& Gillett, F. 1998, PASP, 110, 480
\reference{bl98} Blackwell, D.E., \& Lynas-Gray, A.E. 1998, A\&AS, 129, 505
\reference{co96} Cohen, M., Witteborn, F.C., Carbon, D.F., Davies, J.K., Wooden, D.H., \& Bregman, J.D. 1996, \aj, 112, 2274
\reference{di93} Di Benedetto, G.P. 1993, \aap, 270, 315
\reference{di91} Di Giacomo, A., Lisi, F., Calamai, G., \& Richichi, A. 1991, \aap, 249, 397
\reference{dy96a} Dyck, H.M., Benson, J.A., van Belle, G.T., \& Ridgway, S.T. 1996a, \aj, 111, 1705
\reference{dy96b} Dyck, H.M., van Belle, G.T., \& Benson, J.A. 1996b, \aj, 112, 294
\reference{dy98} Dyck, H.M., van Belle, G.T., \& Thompson, R.R. 1998, \aj, 116, 981
\reference{eg91} Egret, D., Wenger, M., \& Dubois, P. 1991, in Databases and On-line Data in Astronomy, ed. Albrecht, M.A. \& Egret, D. (Dordrecht, Kluwer), p. 79
\reference{ep97} Epchtein, N. 1997, in The Impact of Large Scale Near-infrared Surveys, ed. Garzon, F., Epchtein, N., Omont, A., Burton, W.B., \& Persi, P.(Dordrecht: Kluwer)
\reference{fr88} Fracassini, M., Pasinetti-Fracassini, L.E., Pastori, L., \& Pironi, R. 1988, Bull. D'Inf. Cent. Donnees Stellaires, 35, 121
\reference{ge97} Genova, F., Bonnarel, F., Bartlett, J.G., Dubois, P., Egret, D., Jasniewicz, G. et al. 1997, Baltic Astronomy, 6, 192
\reference{ge97} Gezari, D.Y., Pitts, P.S., \& Schmitz, M. 1997, Catalog of Infrared Observations, Edition 4, unpublished but available online at http://adc.gsfc.nasa.gov/
\reference{ge93} Gezari, D.Y., Schmitz, M., Pitts, P.S., \& Mead, J.M. 1993, Catalog of Infrared Observations, 3rd Edition, NASA Reference Publication 1294
\reference{gi96} Gilliland, R.L., \& Dupree, A.K. 1996, \apj, 463, 29
\reference{gu96} Gunther, J. 1996, JAVSO, 24, 17
\reference{ha74} Hanbury Brown, R., Davis, J., Lake, R.J.W., \& Thompson, R.J. 1974, MNRAS, 167, 475
\reference{hu89} Hutter, D.J., Johnston, K.J., Mozurkewich, D., Simon, R.S., Colavita, M.M., Pan, X.P. et al. 1989, \apj, 340, 1103
\reference{jo68} Johnson, H.L., 1968, in Nebulae and Interstellar
Matter, edited by B.M. Middlehurst and L.H. Aller (University of
Chicago Press, Chicago), Chap. 5
\reference{ka91} Karovska, M., Nisenson, P., Papaliolios, C., \& Boyle, R. P. 1991, \apj, 374, 51
\reference{la97} Lattanzi, M.G., Munari, U., Whitelock, P.A., \& Feast, M.W. 1997, \apj, 485, 328
\reference{ma80} Mathis, J.S., 1980, ARA\&A, 28, 37
\reference{me97} Mermilliod, J.-C., Mermilliod, M., \& Hauck, B. 1997, A\&AS, 124, 349
\reference{mo99} Monnier, J.D., Tuthill, P.G., Lopez, B., Cruzalebes, P., Danchi, W.C., \& Haniff, C.A. 1999, \apj, 512, 351
\reference{mo99} Mozurkewich, D. 1999, private communication
\reference{mo91} Mozurkewich, D., Johnston, K.J., Simon, R.S., Bowers, P. F., Gaume, R., Hutter, D.J. et al. 1991, \aj, 101, 2207
\reference{no99} Nordgren, T. 1999, private communication
\reference{pe93} Percy, J. R., Mattei, J. A. 1993, \apss, 210, 137
\reference{pe98} Perrin, G., Coudé du Foresto, V., Ridgway, S.T., Mariotti, J.-M., Traub, W.A., Carleton, N.P., \& Lacasse, M.G. 1998, \aap, 331, 619
\reference{pe97} Perryman, M.A.C., Lindegren, L., Kovalevsky, J., Hog, E., Bastian, U., Bernacca, P.L. et al. 1997, \aap, 323, L49
\reference{pr92} Press, W.H., Teukolsky, S.A., Vetterling, W.T., \& Flannery, B.P. 1992, Numerical Recipes in C (Port Chester, NY: Cambridge University Press), 689
\reference{ri88} Richichi, A., Salnari, P., \& Lisi, F. 1988, \apj, 326, 791
\reference{ri91} Richichi, A., Lisi, F., \& Calamai, G. 1991, \aap, 241, 131
\reference{ri92} Richichi, A., Lisi, F., \& Di Giacomo, A. 1992, \aap, 254, 149
\reference{ri92} Richichi, A., Di Giacomo, A., Lisi, F., \& Calamai, G. 1992, \aap, 265, 535
\reference{ri95} Richichi, A., Chandrasekhar, T., Lisi, F., Howell, R.R., Meyer, F., Rabbia, Y. et al. 1995, \aap, 301, 439
\reference{ri98a} Richichi, A., Ragland, S., \& Fabbroni, L. 1998a, \aap, 330, 578
\reference{ri98b} Richichi, A., Stecklum, B., Herbst, T.M., Lagage, P.-O., \& Thamm, E. 1998b, \aap, 334, 585
\reference{ri98c} Richichi, A., Ragland, S., Stecklum, B., \& Leinert, C. 1998c, \aap, 338, 527
\reference{ri99} Richichi, A., Fabbroni, L., Ragland, S., \& Scholz, M. 1999, A\&A, 344, 511
\reference{ri74} Ridgway, S.T., Wells, D.C., \& Carbon, D.F. 1974, \aj, 79, 1079
\reference{ri77a} Ridgway, S.T., Wells, D.C., \& Joyce, R.R. 1977, \aj, 82, 414
\reference{ri77b} Ridgway, S.T. 1977, \aj, 82, 511
\reference{ri79} Ridgway, S.T., Wells, D.C., Joyce, R.R., \& Allen, R.G. 1979, \aj, 84, 247
\reference{ri80a} Ridgway, S.T., Jacoby, G.H., Joyce, R.R., \& Wells, D.C. 1980a, \aj, 85, 1496
\reference{ri80b} Ridgway, S.T., Joyce, R.R., White, N.M., \& Wing, R.F., 1980b, \apj, 235, 126
\reference{ri82a} Ridgway, S.T., Jacoby, G.H., Joyce, R.R., Siegel, M.J., \& Wells, D.C. 1982a, \aj, 87, 680
\reference{ri82b} Ridgway, S.T., Jacoby, G.H., Joyce, R.R., Siegel, M.J., \& Wells, D.C. 1982b, \aj, 87, 808
\reference{ri82c} Ridgway, S.T., Jacoby, G.H., Joyce, R.R., Siegel, M.J., \& Wells, D.C. 1982b, \aj, 87, 1044
\reference{sc86} Schmidtke, P.C., Africano, J.L., Jacoby, G.H., Joyce, R.R., \& Ridgway, S.T. 1986, \aj, 91, 961
\reference{tu99} Tuthill, P.G., Haniff, C.A., Baldwin, J.E. 1999,
MNRAS, 306, 353
\reference{va96} van Belle, G.T., Dyck, H.M., Benson, J.A., \& Lacasse, M.G. 1996, \aj, 112, 2147
\reference{va97} van Belle, G.T., Dyck, H.M., Thompson, R.R., Benson, J.A., \& Kannappan, S.J. 1997, \aj, 114, 2150
\reference{va99a} van Belle, G.T., Lane, B.F., Thompson, R.R., Boden, A.F., Colavita, M.M., Dumont, P.J. et al. 1999a, \aj, 117, 521
\reference{va99b} van Belle, G.T., Thompson, R.R., \& Creech-Eakman, M. 1999b in preparation

\end{references}
\end{document}